\documentclass[twocolumn]{aastex631}
\usepackage{float}
\usepackage{amsmath}
\newcommand{\fermi}{\textit{Fermi}}
\newcommand{\gr}{$\gamma$-ray}
\newcommand{\optname}{PKS~2254+074}

\newcommand{\Lagr}{$\mathcal{L}$}

\begin{document}

\title{PKS 2254+074: A Blazar in Likely Association with the Neutrino Event IceCube-190619A}

%%\title{PKS 2254+074 in likely association with IceCube-190619A}

\author{Shunhao Ji}
\affiliation{Department of Astronomy, School of Physics and Astronomy, Yunnan
University, Kunming 650091, China; jishunhao@mail.ynu.edu.cn; wangzx20@ynu.edu.cn}

\author[0000-0003-1984-3852]{Zhongxiang Wang}
\affiliation{Department of Astronomy, School of Physics and Astronomy, Yunnan
University, Kunming 650091, China; jishunhao@mail.ynu.edu.cn; wangzx20@ynu.edu.cn}
\affiliation{Shanghai Astronomical Observatory, Chinese Academy of Sciences, 80
Nandan Road, Shanghai 200030, China}

\begin{abstract}
We report our study of the field of a $\simeq$0.2\,PeV neutrino event 
	IC-190619A. This neutrino belongs to Gold events, which more
	likely have an astrophysical origin. Among the two
	$\gamma$-ray sources within the neutrino's positional uncertainty 
	region, we find
	that one of them, the BL-Lac--type blazar PKS~2254+074, 
	had a $\gamma$-ray flare at the 
	arrival time of the neutrino. The flare is determined to have lasted
	$\sim$2.5\,yr in a 180-day binned light curve, constructed from
	the data collected with the Large Area Telescope (LAT) onboard
	{\it the Fermi Gamma-ray Space Telescope (Fermi)}. Accompanying 
	the flare, optical and mid-infrared brightening is also seen. In 
	addition, $\geq$10\,GeV high energy photons from the source have 
	been detected, suggesting a hardening of the emission during the flare.
	Given both the positional and temporal coincidence of PKS~2254+074
	with IC-190619A, we suggest that this blazar is likely another member
	of a few recently identified (candidate) neutrino-emitting blazars.
\end{abstract}

\keywords{Blazars (164); Gamma-ray sources (633); Neutrino astronomy (1100)}

\section{Introduction} \label{sec:intro}

The IceCube South Pole neutrino observatory \citep{icecube} has been 
detecting neutrino events with energy from $\sim$30\,TeV to PeV
since 2010 \citep{hiIC}. These high energy neutrinos have a high probability
of arising from extraterrestrial sources. While the origin of such events
had been under intense investigation, it was not until 2017 that the association
of such an event, IceCube-170922A, and an extra-galactic source, the blazar 
TXS~0506+056, was established \citep{txs1,txs2}. 
In addition, more recently the detection of the neutrino emission 
from the nearby Seyfert galaxy NGC~1068 at a signficance of 4.2$\sigma$ 
and that from
the Galactic plane at a 4.5$\sigma$ significance level have also been
reported \citep{ngc1068,gal}.

Blazars, the subclass of
Active Galactic Nuclei (AGNs) with a relativistic jet pointing close to 
the line of sight, indeed have been suggested to be neutrino emitters 
(e.g., \citealt{kad+16} and references therein). Protons possibly carried
in a jet would interact with low-energy photons ($p\gamma$ interaction), 
and the produced pions decay and produce neutrinos and high-energy photons.
One notable feature of the TXS~0506+056 case 
is that the blazar was undergoing a \gr\ flare; the peak flux reached 
$\simeq 5\times 10^{-7}$\,ph\,cm$^{-2}$\,s$^{-1}$ in the energy of $>$0.1\,GeV
and the accompanying upto 400\,GeV very-high-energy (VHE) emission was 
also observed \citep{txs1}. 

Following this identification, studies for connecting IceCube-detected 
neutrinos to blazars have been extensively carried out
(e.g., \citealt{ggp+20, fra+20, sta+22, pla+23, rod+24}). Among them,
one is to find a flaring blazar that is positionally coincident
with a reported neutrino event, which would strongly suggest the association
between the two and thus enable the identification of the blazar as being
a neutrino source.
Thus far, the reported cases are PKS B1424$-$418 \citep{kad+16},
GB6~J1040+0617 \citep{gar+19}, MG3~J225517+2409 \citep{fra+20}, GB6~J2113+1121
\citep{lia+22}, PKS~0735+178 \citep{sah+23}, and NVSS~J171822+423948 
\citep{jia+24}. In addition, 
there are two other sources 1H~0323+342 
and 3HSP~J095507.9+355101; the former is a radio-loud narrow-line Seyfert 
galaxy in a minor \gr\ flare \citep{fra+20} and the latter
had an X-ray flare \citep{gpo+20}.

In our examination of neutrinos of high-probability astrophysical
origin (i.e., Gold events that have signalness $\geq$50\%; 
\citealt{abb+23}), we noted a $\sim$0.2 PeV track-like event 
IC-190619A. For its either originally reported positional uncertainty region
\citep{IC190619A}
or that updated in the IceCube Event Catalog of Alert Tracks (IceCat-1;
\citealt{abb+23}), we could include the blazar PKS~2254+074 in it. This
BL-Lac--type blazar has redshift $z = 0.19$ \citep{sfk88,pen+21} and is a
\gr\ source detected with the Large Area Telescope (LAT) onboard {\it the Fermi
Gamma-ray Space Telescope (Fermi)} (e.g., \citealt{dr4}). We also noted that
this source had been in quiescence at $\gamma$-rays in most of the \fermi-LAT
observation, but had a 
flare over the arrival time of 
IC-190619A. Thus, both the positional and temporal coincidence support that
PKS~2254+074 is likely another neutrino-source case. We conducted analysis of 
the archival data for this blazar, and report the results in this paper.
In this work, the cosmological parameters, 
$H_0$ = 67.7\,km\,s$^{-1}$\,Mpc$^{-1}$, $\Omega_m$ = 0.31, 
and $\Omega_\Lambda$ =0.69,
from the Planck mission \citep{Planck} were used. 

\section{Data Analysis and Results}\label{data}

\subsection{Positional coincidence analysis}

\begin{figure}[!ht]
\centering
\includegraphics[width=0.49\textwidth]{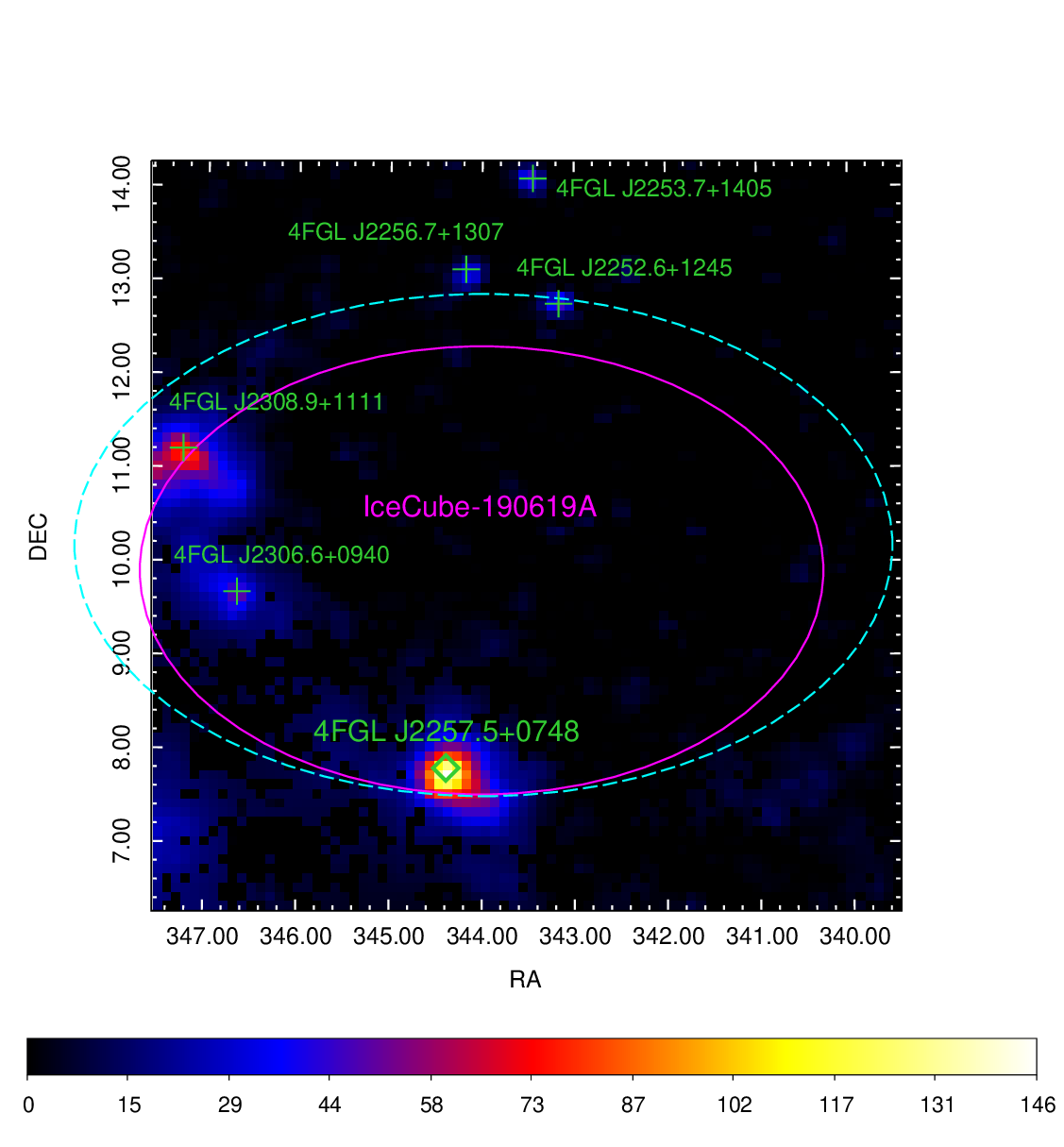}
\caption{TS map of a size of 8\degr $\times$ 8\degr\ centered at IC-190619A
	in the energy of 0.1--500\,GeV. 
	The dashed cyan ellipse marks the uncertainty region with the 
	systematic uncertainty (by scaling the major and minor 
	axis of the 90\% error ellipse with a factor of 1.3; \citealt{ggp+20})
	included,
and the pink ellipse marks the uncertainty region given in IceCat-1. 
	In addition to PKS 2254+074 
	(4FGL J2257.5+0748), another \fermi\ LAT source 
	(4FGL~J2306.6+0940) is also in the uncertainty region in IceCat-1.
\label{fig:tsmap}}
\end{figure}

On 2019 June 19 13:14:18.04 UT (MJD 58653.55), IceCube detected 
a $\sim$0.2 PeV track-like event IC-190619A with signalness $\simeq$ 54.6\%\footnote{\url{https://gcn.gsfc.nasa.gov/notices_amon_g_b/132707_54984442.amon}}.
The initial automated alert provided 
the arrival direction,
%R.A. = 343\degr.26, Decl. = +10\degr.73 (equinox J2000.0), with a 90\% error radius of 2.71 deg. 
which was subsequently updated from the more sophisticated
reconstruction algorithms to be 
R.A. = 343\degr.26$_{-2.63}^{+4.08}$, Decl. = +10\degr.73$_{-2.61}^{+1.51}$ 
(equinox J2000.0; \citealt{IC190619A}).  
The blazar PKS 2254+074 (4FGL J2257.5+0748) is slightly outside the uncertainty 
region. However, when we added the systematic uncertainty often considered
(e.g., \citealt{ggp+20,Plavin+20,Hovatta+21,sah+23}),
the source is included in the combined uncertainty region
(Figure~\ref{fig:tsmap}). Furthermore in IceCat-1, the neutrino's position 
was updated to be R.A. = 343\degr.52$_{-3.16}^{+4.13}$, 
Decl. = +10\degr.28$_{-2.76}^{+2.02}$, having the blazar well enclosed 
(Figure~\ref{fig:tsmap}).

No \gr\ sources in the LAT 8-year Source Catalog (at the time) were within 
the initial 90\% statistical uncertainty region \citep{IC190619A}.
However in the latest
\fermi\ LAT Source Catalog (i.e., 4FGL-DR4; \citealt{dr4}), there is
4FGL J2306.6+0940, classified as blazar of uncertain type, also in 
the uncertainty region in IceCat-1. 
This source, as well as several other sources near but outside the error 
region (Figure~\ref{fig:tsmap}),
was fainter than 4FGL J2257.5+0748. The test statistic (TS) value 
for it was 48, which corresponds to
the detection significance of 6.9$\sigma$ ($\simeq \sqrt{TS}$). We checked 
its $\gamma$-ray light curve, and found it did not show any significant 
variations at the neutrino's arrival time.

\subsection{{\it Fermi}-LAT data analysis}

\subsubsection{Data and source model}
The {\it Fermi}-LAT data used were 0.1--500 GeV photon events (evclass=128 
and evtype=3) from the updated \fermi\ Pass 8 database in a time range of 
from 2008-08-04 15:43:36 (UTC) to 2024-04-18 00:05:53 (UTC). The region of 
interest (RoI) was set to be 20\degr $\times$ 20\degr\ centered 
at PKS 2254+074. We excluded the events with zenith 
angles $>$ 90\degr\ to avoid the Earth-limb contamination. The expression 
DATA\_QUAL $>$ 0 \&\& LAT\_CONFIG = 1 was applied to select good time-interval 
events. 
%In our analysis, the package Fermitools–2.2.0 and the instrumental response function P8R3\_SOURCE\_V3 were used.

\begin{figure*}[!ht]
\centering
\includegraphics[width=0.49\textwidth]{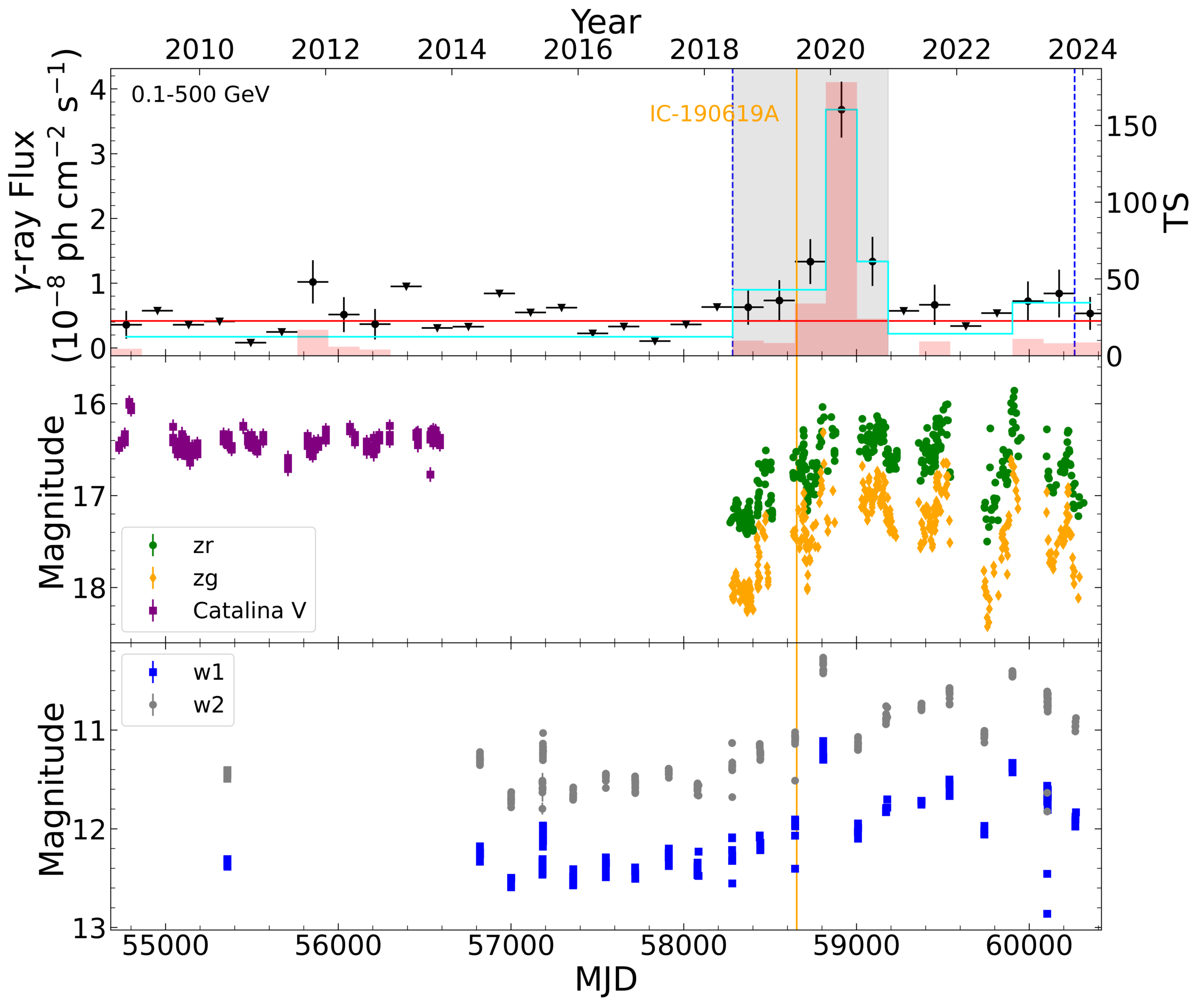}
\includegraphics[width=0.49\textwidth]{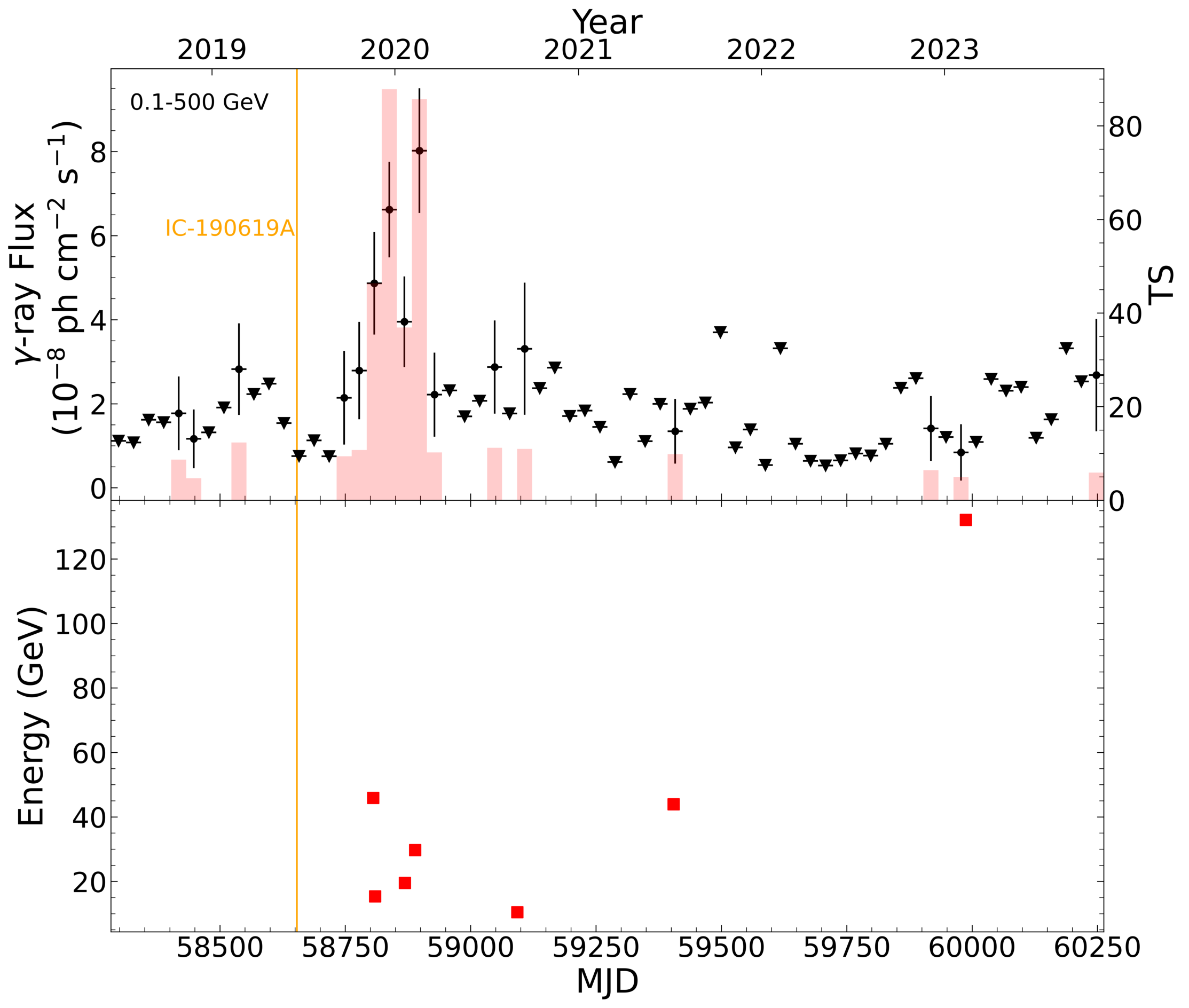}
	\caption{{\it Left:} 180-day binned $\gamma$-ray light curve of
	PKS~2254+074 in 0.1--500 GeV (upper), its
	optical Catalina $V$ band and ZTF
	$zg$ and $zr$ band light curves (middle), and MIR WISE light curves
	(bottom). The orange line marks the arrival time 
	of IC-190619A. In the top panel, the red line represents 
	the average flux, the cyan histograms mark the blocks,
and the gray region marks the flare defined by the HOP algorithm. {\it Right:}
	30-day binned $\gamma$-ray light curve during the time period marked 
	by the two blue dashed lines in the left panel (upper) and
	the arrival times of $\geq$10\,GeV photons with probability $>$ 68\% 
	(lower; see Section~\ref{spec}). 
	In both \gr\ light curve panels, the downward triangles
	are the 95\% C.L. flux upper limits 
	and red histograms indicate the TS values of the data points.
	\label{fig:lc}}
\end{figure*}

The source model was generated based on 4FGL-DR4. 
In the catalog, PKS 2254+074 (4FGL J2257.5+0748) was modeled as a point source 
with a power-law (PL) spectrum, $dN/dE = N{_0}(E/E{_0})^{-\Gamma}$, 
where $E{_0}$ was 
fixed at 1.55 GeV. We adopted this spectral model.
All other sources in 4FGL-DR4 within 25\degr\ of the target were included
in the source model.
The spectral indices and normalizations of the sources within 5\degr\ 
of the target were set as free parameters and the other parameters were
fixed at the 
catalog values. The extragalactic diffuse emission and 
the Galactic diffuse emission components,  
iso\_P8R3\_SOURCE\_V3\_v1.txt and gll\_iem\_v07.fits respectively, were
also included.
Their normalizations were always set as free parameters in our analysis.

\subsubsection{Light-curve analysis}\label{lc_ana}

Using the source model described above, we performed the standard binned 
likelihood analysis to the whole data in 0.1--500 GeV for \optname. 
We obtained photon index $\Gamma$ = 2.18$\pm$0.08, with a TS value of 136 
($\simeq$11.7$\sigma$ detection signficance). 
The results are consistent with those given in 4FGL-DR4.

We extracted the 0.1--500 GeV $\gamma$-ray light curve of the source
by setting a 180-day time bin and performing the maximum likelihood analysis 
to the binned data. In the extraction, only the normalization parameters of 
the sources within 5\degr\ of the target were set free and the other 
parameters were fixed at the best-fit values obtained in the analysis of 
the whole data. For the data points with TS$<$4, we computed the 95\% 
confidence level (C.L.) upper limits.
As revealed by the light curve (left panel of 
Figure \ref{fig:lc}), PKS 2254+074 had been nearly not detectable
for $\sim$10\,yr. Since 2018, a $\gamma$-ray flare peaking 
at $\sim$MJD~58900 was observed, during which IC-190619A was 
detected. To define the time duration of the flare, we 
employed the Bayesian block algorithm \citep{Scargle+13} and HOP 
algorithm \citep{Meyer+19}, implemented through a python 
code\footnote{\url{https://github.com/swagner-astro/lightcurves}} \citep{Wagner+22}. 
The start/end of a segment (e.g., a flare) is determined by the
flux of a block exceeding above or dropping under the average flux
(see details and the so-called HOP algorithm in \citealt{Meyer+19}).
The duration determined is from MJD~58282.66 to 59182.66 ($\sim$2.5\,yr),
with the peak of the flare (the middle time 
of the highest-flux data point) being MJD 58912.66.
For the time ranges excluding the flare, we defined them as the low state
of the source.

We also constructed the optical and mid-infrared (MIR) light curves
of PKS~2254+074 from the data taken respectively from the
the Catalina Real-Time Transient Survey \citep{ddm+09}, the
Zwicky Transient Facility (ZTF; \citealt{Bellm+19}), and the
NEOWISE Single-exposure Source Database \citep{Mainzer+14}.
The optical bands are Catalina $V$, ZTF $g$ and $r$ 
(named as $zg$ and $zr$ 
respectively), and the MIR are WISE w1 (3.4\,$\mu$m) and w2 
(4.6\,$\mu$m).  The light curves show that the optical and MIR emissions 
accompanied the \gr\ flare and
brightened by $\sim$1\,mag. During the flare and afterwards,
strong variations, with magnitude changes as large as $\sim$1, are also
seen.

To show more details of the $\gamma$-ray flare, we further extracted 
a 30-day binned light curve from MJD~58282.66 to 60262.66
(P1; the region between the two blue dashed lines in the left panel of 
Figure~\ref{fig:lc}); the same extraction process as the above
was conducted. The light curve (right panel of 
Figure~\ref{fig:lc}) further reveals a strong flaring activity, 
which lasted $\sim$200\,d
from $\sim$MJD~58750 to 58950. During the time period, the peak flux reached
$\sim 8\times 10^{-8}$\,ph\,cm$^{-2}$\,s$^{-1}$.

\subsubsection{Spectrum analysis}\label{spec}

We performed the likelihood analysis to the data of the two states 
(i.e., the flare and and low state)
in 0.1--500 GeV. For the flare, the obtained best-fit 
$\Gamma$ = 2.06$\pm$0.07 and photon flux 
$F_{\gamma}$ = (1.21$\pm$0.20) $\times$ 10$^{-8}$\,ph\,cm$^{-2}$\,s$^{-1}$
(TS $\simeq$ 190 or detection significance $\simeq$ 13.8$\sigma$). For the low state, $\Gamma$ = 2.24$\pm$0.15 and 
$F_{\gamma}$ = (2.50$\pm$0.94) $\times$ 10$^{-9}$\,ph\,cm$^{-2}$\,s$^{-1}$
(TS$\simeq$38 or detection significance $\simeq$ 6.2$\sigma$). Because of the large uncertainties in the latter,
no spectral changes between the two states could be determined.

We then obtained the $\gamma$-ray spectrum of PKS 2254+074 in the flare by
performing the binned likelihood analysis to the data in 8 evenly divided 
energy bins in logarithm from 0.1 to 500 GeV. In this analysis, the 
spectral normalizations of the sources in the source model within 5\degr\ of 
the target were set free and all other spectral parameters of the sources 
were fixed at the values obtained in the above likelihood analysis to the data 
of the flare. For the spectrum, we only kept the data points with TS $\geq$ 4. 

For the low state, no decent spectrum can be obtained because of the low
detection significance (i.e., TS $\simeq$ 38).

\begin{figure}[!ht]
\centering
\includegraphics[width=0.47\textwidth]{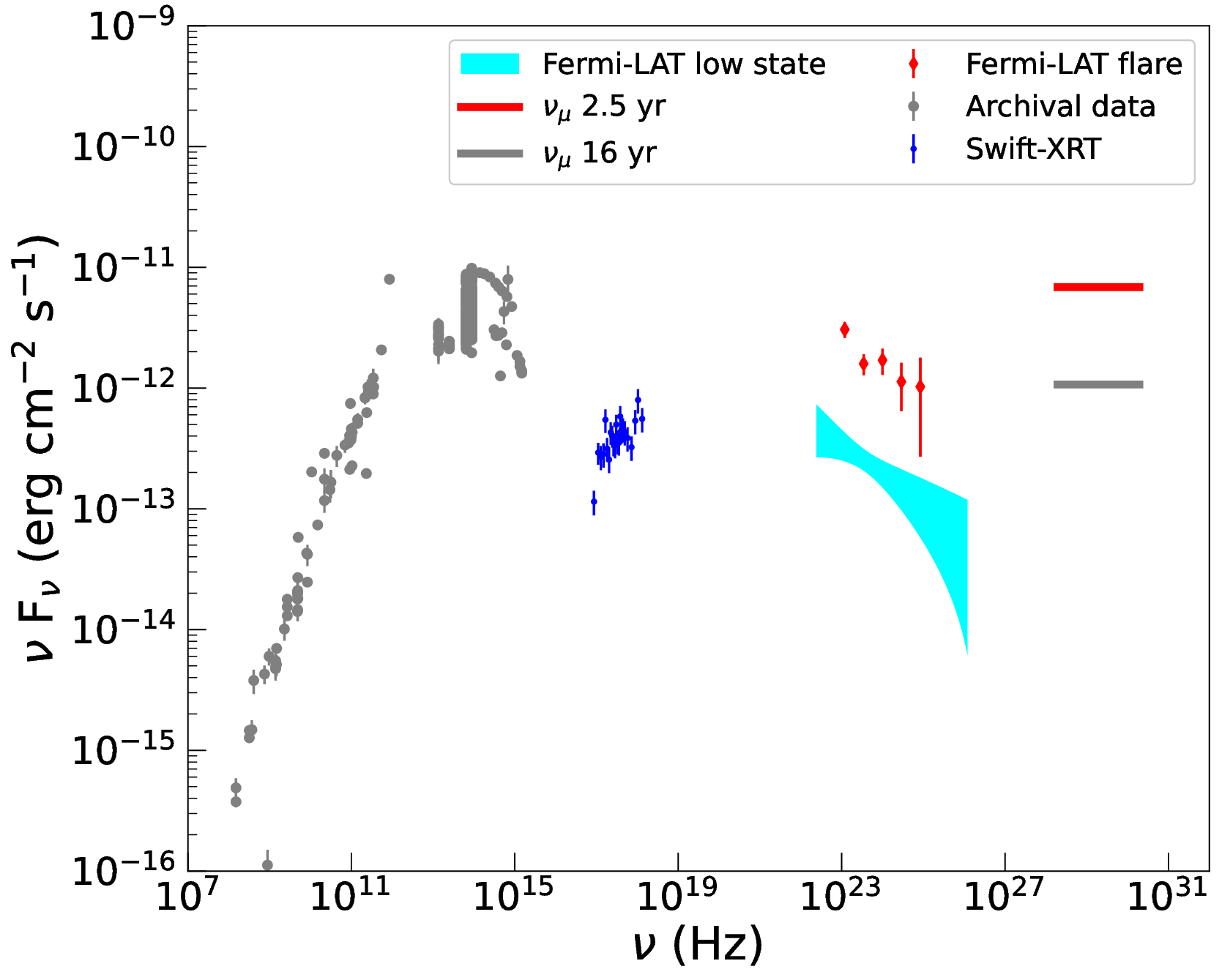}
\caption{Broadband SED of PKS 2254+074. The gray dots are the archival data 
	from Firmamento. The Swift-XRT X-ray spectrum is 
	plotted as the blue dots. The $\gamma$-ray spectrum in the flare
	and the model fit in the low state are plotted as red diamonds
	and cyan region, respectively. The estimated neutrino fluxes for
	two time intervals (see Section~\ref{dis}) are shown as the red and 
	gray lines.
	\label{fig:sed}}
\end{figure}

Several blazars showed spectrum hardening during the flares that were
temporally coincident
with the neutrinos' arrival times (e.g., \citealt{gar+19,gpo+20,lia+22}), 
including
TXS~0506+056 that emitted VHE photons \citep{txs1}. We thus checked the high
energy, $\geq$10\,GeV photons from PKS 2254+074.
Running {\tt gtsrcprob} on the whole data in an 1\degr\ RoI with
the best-fit model from the analysis of the whole data,
we found five such photons with $>$68\% probabilities.
They were mostly during the flare. However, one had 132\,GeV 
the highest energy with 95\% probability, arrived after the flare 
on MJD 59987.38.
The model used for the target can affect the estimation of photon probability. 
When using the best-fit model from the analysis of the P1 data
(MJD 58282.66--60262.66), we found seven high-energy photons, with 
probabilities $\geq$77\%. The arrival times and energies of these photons
are shown in the bottom right panel of Figure~\ref{fig:lc}.

\subsection{Broadband spectral energy distribution}\label{sed}

We constructed the broadband spectral energy distribution (SED) of 
PKS 2254+074 from radio to $\gamma$-ray band. The archival radio to optical 
data were obtained using the tool Firmamento\footnote{\url{https://firmamento.hosting.nyu.edu/data_access}}. 
The X-ray spectral data points were extracted from the observations conducted
with the X-ray Telescope (XRT) onboard {\it the Neil Gehrels Swift Observatory
(Swift)}. Details of the X-ray data and analysis are described in Appendix
Section~\ref{xrt}. 
Both the \gr\ spectrum in the flare and the model fit in the low state
are included in the SED (Figure \ref{fig:sed}).

\section{Discussion}
\label{dis}

Our analysis has found another blazar, PKS 2254+074, as a possible
neutrino source. Its position is within the uncertainty region of
the IceCube Gold event IC-190619A, and it had a flare temporally coincident
with the arrival time of the neutrino. Such matches have established the likely
association of a few blazars with neutrinos. In addition, we have detected
several high-energy photons during the flare, which had not been seen in
the source's long-term emission, suggesting the hardening of the source's 
emission in the flare. This feature could be another similarity \citep{lia+22}.
We note that the peak flux of this blazar's flare reached 
$\sim 8\times 10^{-8}$\,ph\,cm$^{-2}$\,s$^{-1}$, which is also comparable with
the values of those (candidate) neutrino blazars \citep{lia+22}. 
These similarities strongly support the possibility of PKS 2254+074 
as another
member of the neutrino blazars.

PKS 2254+074 is classified as a low-synchrotron
peaked (LSP) blazar in \citet{Ajello+22}, and its
peak frequency 
$\nu_{\rm pk}^{\rm syn}\simeq 4.0_{-2.0}^{+3.9}\times 10^{13}$\,Hz
(given by Firmamento using a machine learning estimator BLAST).
Its long-term average \gr\ luminosity \Lagr$_{\gamma}$ and $\Gamma$ we obtained 
are $3.7 \times 10^{44}$\,erg\,s$^{-1}$ and 2.18, 
and during the (2.5\,yr) flare, the values are 
$1.5 \times 10 ^{45}$\,erg\,s$^{-1}$ and 2.06, respectively. 
Comparing it to the previously identified neutrino blazar candidates,
it has the lowest redshift and the lowest long-term luminosity. The other 
neutrino blazars are generally in a redshift range of 0.3--1.5, but
the very recently discovered one NVSS~J171822+423948 has $z = 2.68$
\citep{jia+24}. Their \gr\ luminosities
are in a range of $\sim 4\times 10^{45}$--$5\times 10^{48}$. 
Thus, if it is truely a neutrino
source, the detection could be due to its relatively close distance.
The photon index and luminosity values of PKS 2254+074 are typical 
for a BL Lac (e.g., \citealt{Ackermann+15,Chen+18}). 
Among the other neutrino blazars, GB6 J1040+0617 is in the same class,
a BL Lac LSP. PKS B1424$-$418, GB6 J2113+1121, and NVSS J171822+423948 
are also LSPs although they are flat-spectrum radio quasars (FSRQs).
TXS 0506+056 and the other candidate 3HSP J095507.9+355101, which
are an intermediate- and a high-synchrotron peaked blazars, respectively,
have previously been noted as the outliers of the so-called 
blazar sequence \citep{gpo+20}. 
We also note that if we take the average \Lagr$_{\gamma}$ 
and $\nu_{\rm pk}^{\rm syn}$
values of PKS 2254+074, it could also be an outlier
(because its \Lagr$_{\gamma}$ $<10^{45}$\,erg\,s$^{-1}$
while $\nu_{\rm pk}^{\rm syn}\simeq 4.0\times 10^{13}$\,Hz;
see Figure 4 in \citealt{gpo+20}). Unfortunately,
we do not have information for its $\nu_{\rm pk}^{\rm syn}$ in the flare.
Whether its flaring variations are similar to those of 
TXS 0506+056 and 3HSP J095507.9+355101 are not known.

Following \citet{gpo+20} and \citet{jia+24}, we estimated the neutrino flux
given the detection of 1 muon neutrino event. Considering the effective
area $A_{\text{eff}}\simeq 24$ m$^{2}$ (GFU\_Gold, see \citealt{abb+23}),
a PL neutrino energy spectrum with an index of $-$2 from 80\,TeV to 
8\,PeV \citep{Oikonomou+21}, the integrated muon neutrino energy flux 
would be 3.2$\times$10$^{-11}$\,erg\,cm$^{-2}$\,s$^{-1}$ and 
4.9$\times$10$^{-12}$\,erg\,cm$^{-2}$\,s$^{-1}$ (the corresponding differential
spectra are shown in Figure~\ref{fig:sed}), while
the two values correspond to
the neutrino-emitting time intervals of 2.5\,yr (flare time duration) 
and 16\,yr (approximate time length of the \fermi-LAT data), respectively. 
The corresponding luminosities \Lagr$_{\nu_{\mu}}$ would be
3.4$\times$10$^{45}$\,erg\,s$^{-1}$ and 5.4$\times$10$^{44}$\,erg\,s$^{-1}$,
and the luminosity term ${\epsilon_{\nu_{\mu}}}L_{\nu_{\mu}}$
(see \citealt{gpo+20}),
approximated with
${\epsilon_{\nu_{\mu}}}L_{\nu_{\mu}}$ $\sim$ \Lagr$_{\nu_{\mu}}/ln(8\ {\rm PeV}/80\ {\rm TeV})$, 
would be 7.5$\times$10$^{44}$\,erg\,s$^{-1}$ and 
1.2$\times$10$^{44}$\,erg\,s$^{-1}$, respectively.

The neutrino flux, on the other hand, may be estimated from the observed 
\gr\ flux
(see \citealt{gpo+20} and \citealt{jia+24} for detailed calculations). 
The connection between them is given as \citep{Murase+18}
\begin{equation}\label{ppi}
\begin{split}
    {\epsilon_\nu}{L_{\epsilon_{\nu}}} &\approx \frac{6\left(1+Y_{IC}\right)}{5}{\epsilon_\gamma}{L_{\epsilon_{\gamma}}}\vert_{\epsilon_{\text{syn}}^{p\pi}} \\
    &\approx 8 \times 10^{44} \, \text{erg s}^{-1} \left(\frac{{\epsilon_\gamma}{L_{\epsilon_{\gamma}}}\vert_{\epsilon_{\text{syn}}^{p\pi}}}{7 \times 10^{44}}\right).
\end{split}
\end{equation}
Here, $Y_{IC}$ is the Compton-Y parameter, typically $\leq 1$ \citep{Murase+18}.
Basically during the photopion (${p\pi}$) process, three-eighths of the
proton energy is taken away by all-flavour neutrinos and the remaining energy 
goes to the production of electrons and pionic $\gamma$-rays.
Following \citet{gpo+20}, we estimate the \gr\ luminosity term
$\epsilon_{\gamma}L_{\gamma}$ $\sim$ \Lagr$_{\gamma}/\ln(500\ {\rm GeV}/100\ {\rm MeV})$
$\simeq$1.7$\times$10$^{44}$\,erg\,s$^{-1}$ (for 2.5 yr) or 
$\simeq$4.4$\times$10$^{43}$\,erg\,s$^{-1}$ (for 16 yr). 
Using Eq. \ref{ppi}, the muon neutrino luminosities 
(i.e., $\frac{1}{3}{\epsilon_\nu}{L_{\epsilon_{\nu}}}$) would be 
6.7$\times$10$^{43}$\,erg\,s$^{-1}$ (2.5 yr) or 
1.7$\times$10$^{43}$\,erg\,s$^{-1}$ (16 yr).
Comparing the values with the above obtained by considering the detection of 
one muon neutrino with the IceCube, they are factors of 11 and 7, respectively, 
lower.  The factors are approximately 4--14 times smaller than those 
estimated in the cases reported
by \citet{gpo+20} and \citet{jia+24}, but still indicate a low 
neutrino detection probability (i.e., the Poisson probability to detect one 
neutrino is $\sim$0.1).

\begin{acknowledgments}

This work was based on observations obtained with the Samuel Oschin Telescope 
48-inch and the 60-inch Telescope at the Palomar Observatory as part of the 
Zwicky Transient Facility project. ZTF is supported by the National Science
Foundation under Grant No. AST-2034437 and a collaboration including Caltech, 
IPAC, the Weizmann Institute for Science, the Oskar Klein Center at Stockholm 
University, the University of Maryland, Deutsches Elektronen-Synchrotron
and Humboldt University, the TANGO Consortium of Taiwan, the University of 
Wisconsin at Milwaukee, Trinity College Dublin, Lawrence Livermore National 
Laboratories, and IN2P3, France. Operations are conducted by COO, IPAC, and UW.

This work made use of data products from the Wide-field Infrared Survey 
Explorer, which is a joint project of the University of California, 
Los Angeles, and the Jet Propulsion Laboratory/California Institute of 
Technology, funded by the National Aeronautics and Space Administration.

	We thank the referees for very detailed and helpful comments.
This research is supported by the Basic Research Program of Yunnan Province 
(No. 202201AS070005), the National Natural Science Foundation of China 
(12273033), and the Original Innovation Program of the Chinese Academy of 
Sciences (E085021002). S.J. acknowledges the support of the science research program for graduate students of Yunnan University (KC-23234629).
\end{acknowledgments}

\bibliographystyle{aasjournal}
\bibliography{neu}{}

\appendix
%%\label{sec:app}

\restartappendixnumbering
\section{X-ray Data Analysis}
\label{xrt}
The X-ray data were from the observations conducted with Swift-XRT.
There are six observations between 2007 and 2012, and the information
is given in Table~\ref{tab:xrt}.
We used the online Swift-XRT data products generator 
tool\footnote{\url{https://www.swift.ac.uk/user_objects/}} (for details 
about the online tool, see \citealt{Evans+09}) to extract the 0.3--10 keV 
spectrum of PKS 2254+074 from the six sets of the data. We grouped the spectrum 
to a minimum of 20 counts 
per bin using the GRPPHA task of FTOOLS. We fitted the spectrum with an 
absorbed PL model in the XSPEC 12.12.1, where the Galactic hydrogen column 
density $N_{\rm H}$ was fixed 
at 4.76 $\times$ 10$^{20}$ cm$^{-2}$ \citep{HI4PI..16}. The obtained PL
index was 1.87$\pm$0.1 and unabsorbed flux was 
1.36$\pm0.12\times 10^{-12}$\,erg\,cm$^{-2}$\,s$^{-1}$.
\begin{table}[!ht]
	\centering
        \caption{Information for Swift-XRT observations}
        \label{tab:xrt}
        \begin{tabular}{lcc}
        \hline
        \hline
Date & Obsid & Exposure\\
		&  & (ks) \\\hline
		2007-04-25  & 00036360001 &    5.01\\
		2007-04-28  & 00036360002 &     4.41\\
		2009-01-18  & 00036360003 &     4.02\\
		2009-01-22  & 00036360004 &     2.06\\
		2012-04-24  & 00036360005 &     0.38\\
		2012-07-05  & 00036360006 &     0.63\\
\hline
\end{tabular}
%\tablecomments{}
\end{table}

\end{document}